# Using Virtual Observatory with python: querying remote astronomical databases

**Frédéric Paletou, Ivan Zolotukhin** *(Observatoire Midi-Pyrénées, IRAP, Université Paul Sabatier - Toulouse, France)*

*This basic example is devoted to extending an existing catalogue with data taken elsewhere, either from CDS VizieR or Simbad database. We use the so-called "Spectroscopic Survey of Stars in the Solar Neighborhood" (aka. $S^4N$, Allende Prieto et al. 2004; referenced as* `J/A+A/420/183` *in VizieR) in order to retrieve all objects with available data for the main set of stellar parameters: effective temperature (Teff), surface gravity (logg) and metallicity ([Fe/H]). Then for each object in this dataset we query Simbad database to retrieve the projected rotational velocity (vsini). This combines Vizier and Simbad queries made using python's* `astroquery` *module. Compiling this collection of data is important for many problems in stellar astrophysics, such as the test of a stellar parameter inversion method. The tutorial covers remote database access, filtering tables with arbitrary criteria, creating and writing your own tables and basics of plotting in python.*

**1. Initialize the Python environment**

You can follow this tutorial step by step in any python shell, such as `python` or, more conveniently, `ipython`. Just copy and paste the code snippets below in your python session in the terminal. Note that you have to have matplotlib, numpy, astropy and astroquery python modules installed in your system. Do not hesitate to print any variable below (`print varname`) to explore data structures we use in the tutorial.

```python
import os

#--- import pylab stuff
import pylab as plt
import numpy as np

#--- import astropy stuff
from astropy.table import Table, Column

#--- import astroquery stuff
from astroquery.vizier import Vizier
from astroquery.simbad import Simbad
```

## 2. Proceed with the query of CDS Vizier service

The S⁴N catalogue contains 4 different tables, which command `find_catalogs` and `get_catalogs` allow to identify. We are interested in "table2" which is `s4n[0]` hereafter.

One "trick" is to know name conventions (e.g., `__Fe_H_`) in order to retrieve specific stellar parameters. It may also be useful to get information on units into which they have been expressed.

```python
# this is the trick: tell to return column name 'all'
v = Vizier(columns=['all'])
catalist = Vizier.find_catalogs('J/A+A/420/183')

#--- trick: in order to get all rows in tables
Vizier.ROW_LIMIT = -1
s4n = Vizier.get_catalogs(catalist.keys())
```

At this stage, one has retrieved the content of the S⁴N Vizier catalogue for these stellar fundamental parameters. Notice that some objects do not have the full set of these 3 stellar fundamental parameters available.

Afterwards, we shall keep only these objects for which all of these 3 parameters are set in the S⁴N catalogue.

## 3. Filter data

```python
#--- retain only targets for which all 3 parameters have been set
filter = (s4n[0]['Teff'].mask == False) & (s4n[0]['logg'].mask == False) & \
   (s4n[0]['__Fe_H_'].mask == False)

#--- compute number of targets
ntargets = len(s4n[0][filter])

#--- we shall need to restore full HIP # for Simbad queries...
objhip = []

#--- apply filter to tables
obj  = s4n[0]['HIP'][filter]
teff = s4n[0]['Teff'][filter]
logg = s4n[0]['logg'][filter]
mtal = s4n[0]['__Fe_H_'][filter]
```

**A note on units:** the S⁴N survey expresses metallicities in "H=12" and we convert them in more usual "dex" so that the solar [Fe/H]=0.

```
#--- now convert to usual (dex) metallicity
mtal = mtal - 7.55
```

## 4. Make CDS Simbad queries

For each row in our filtered table derived from the S⁴N catalogue we make a Simbad query by object name (e.g. 'HIP 171') and save `vsini` value in a new array.

```
#--- restoring full HIP names for next (simbad) queries
for i in np.arange(ntargets):
    objhip.append('HIP' + str(obj[i]))

#--- now retrieving vsini value (most recent though) from Simbad
Simbad.add_votable_fields('rot')
vsini=np.zeros((ntargets), 'Float64')

for i in np.arange(1, ntargets):
    t = Simbad.query_object(objhip[i])
    if (len(t) <> 0):
        vsini[i] = t['ROT_Vsini'][0]
    else:
        vsini[i] = -1.
```

*We start the last loop above from index 1 to make sure we avoid querying Simbad with HIP0==Sun in the S⁴N catalogue.*

## 5. Saving result catalogue/table

This can be made using the `astropy.table` module.

```
#--- make my own table...
MyS4N = Table( [objhip, teff, logg, mtal, vsini],
names=('ObjHIP','Teff','logg','[Fe/H]','vsini'))

#--- save table as Fits file
MyS4N.write('MyS4N.fits')
```

This table may be exported in various formats (e.g. VOTable or ASCII) for the record or for data exchange between collaborators (see the `astropy` [documentation](#)). Here we save data as a FITS file.

## 6. A graphical summary of the output

```python
#--- graphical summary/output
plt.figure(1)
plt.scatter(teff, logg, s=50*(mtal+1), c=mtal)
plt.title('S4N (Allende Prieto et al. 2004)', fontsize=20)
plt.xlabel('effective temperature [K]', fontsize=20)
plt.ylabel('surface gravity [dex]', fontsize=20)
cb = plt.colorbar()
cb.set_label('[Fe/H] metallicity', size=20)

plt.figure(2)
plt.scatter(teff, logg, s=vsini, c=vsini)
plt.title('S4N (Allende Prieto et al. 2004)', fontsize=20)
plt.xlabel('effective temperature [K]', fontsize=20)
plt.ylabel('surface gravity [dex]', fontsize=20)
cb = plt.colorbar()
cb.set_label('vsini (Simbad)', size=20)
plt.show()
```

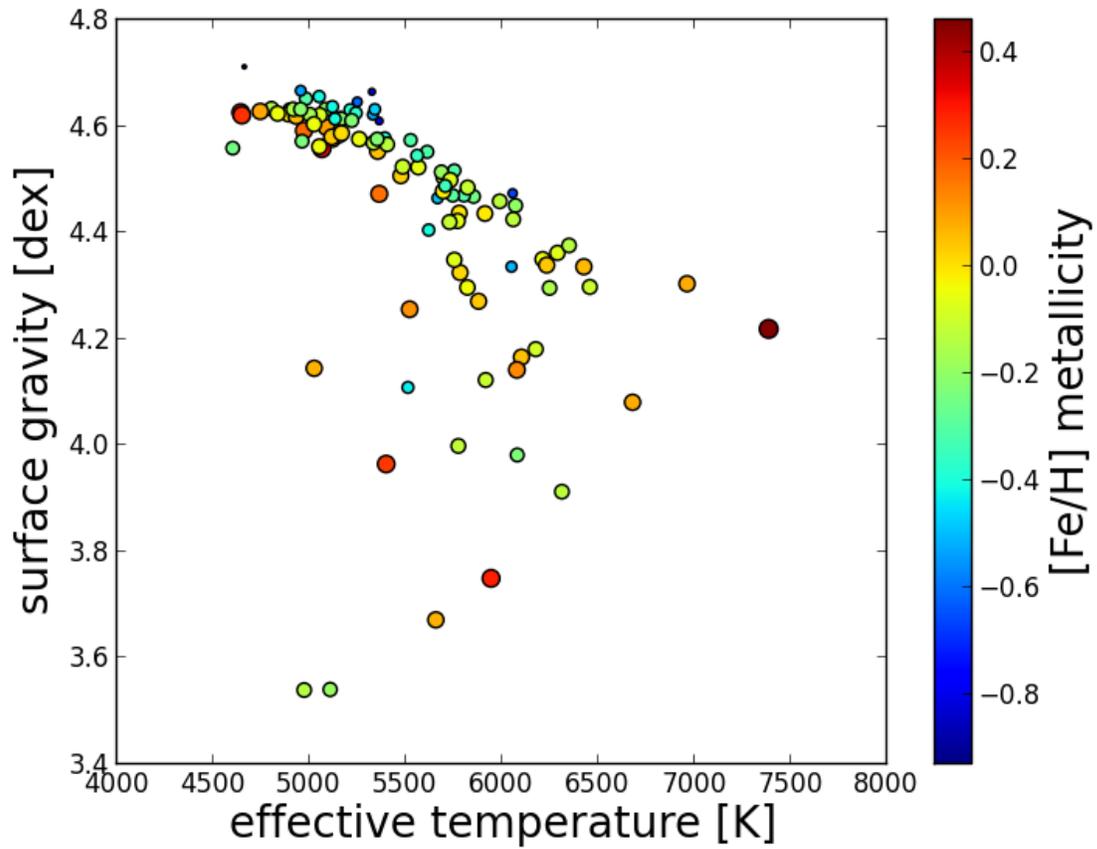

**Fig. 1:** *A summary of collected data from the S⁴N survey. Only data for which {Teff., logg, [Fe/H]} are simultaneously available have been considered. Metallicities have been converted to more usual "dex" values such as the solar [Fe/H]=0.*

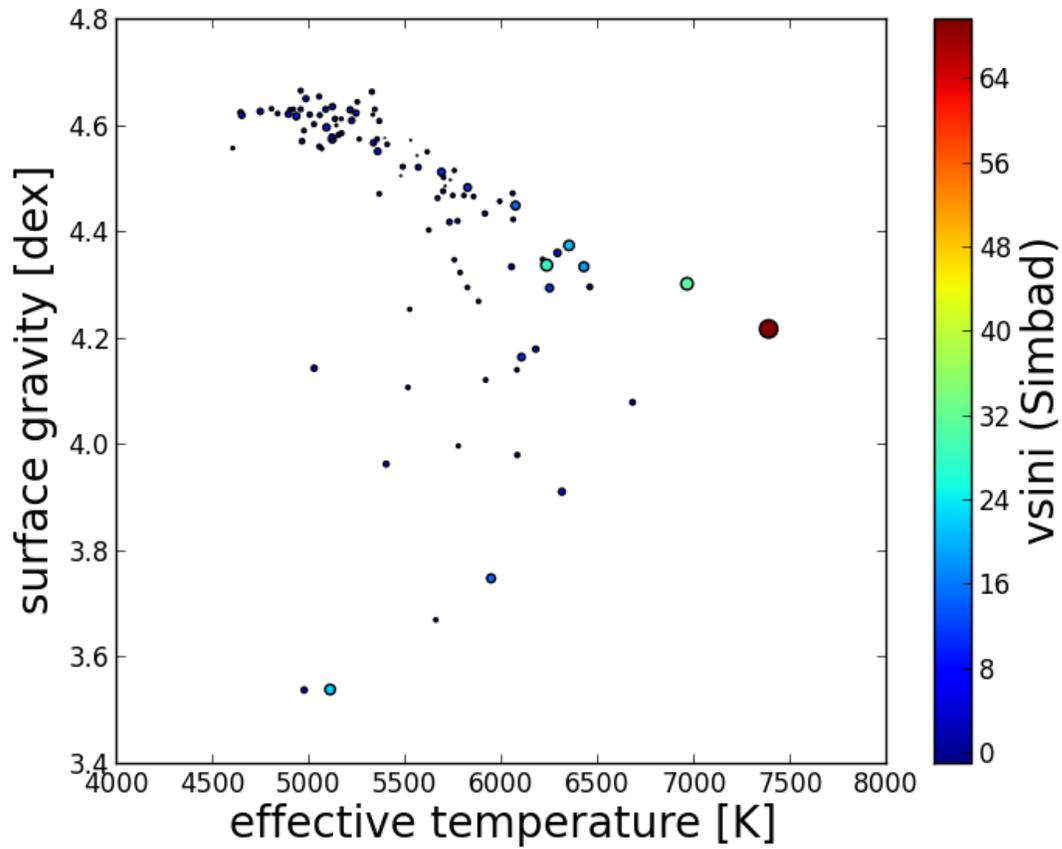

**Fig. 2:** *Same as Fig. 1 but surface/color of the dots now vary with the projected rotational velocity vsini value retrieved from Simbad.*

**You can download the full Python script from this tutorial here:** http://goo.gl/LmdPOI